\def\rn{}
\def\nn#1 #2{#2. #1}				
\def\nnn#1 #2 #3{#2. #3. #1}			
\def\nnnn#1 #2 #3 #4{#2. #3. #4 #1}		
\def\nnnnn#1 #2 #3 #4 #5{#2. #3. #4 #5. #1}	
\def\dualand{ and\hbox{ }}				
\def\multiand{, and\hbox{ }}				
\def\rf#1;#2;#3;#4;#5 {{\frenchspacing\par\rn#1, #3 {\bf #4}, #5 (#2). \par}}
\def\rg#1;#2;#3;#4;#5;#6 {{\frenchspacing\par\rn#1, #3 {\bf #4}, #5 (#2). \par}}
\def\rfbook#1;#2;#3;#4;#5 {{\frenchspacing\par\rn#1, {\it #3} (#5, #4, #2).\par}}
\def\rfprep#1;#2;#3 {{\par\frenchspacing\rn#1, #3 (#2).\par}}
\def\rfproc#1;#2;#3;#4;#5;#6 {{\frenchspacing\par\rn#1 #2, in {\it #3}, ed. #4 (#5: #6)\par}}
\def\rfprocp#1;#2;#3;#4;#5;#6;#7 {{\frenchspacing\par\rn#1 #2, in {\it #3}, ed. #4 (#5: #6), p#7\par}}

\def\rg#1;#2;#3;#4;#5;#6 {\par\rn#1 #2, {\it #3}, {\bf #4}, #5 (``#6'') \par}
\def\rf#1;#2;#3;#4;#5 {\par\rn#1, {\it #3}, {\bf #4}, #5 (#2)\par}
\def\rfbook#1;#2;#3;#4;#5 {{\frenchspacing\par\rn#1, {\it #3} (#4: #5, #2)\par}}
\def\rfproc#1;#2;#3;#4;#5;#6 {{\frenchspacing\par\rn#1 #2, in {\it #3}, ed. #4 (#5: #6)\par}}
\def\rfprocp#1;#2;#3;#4;#5;#6;#7 {{\frenchspacing\par\rn#1 #2, in {\it #3}, ed. #4 (#5: #6), p#7\par}}
\def\rfprep#1;#2;#3  {{\par\rn#1, #3 (#2)\par}}
\def\rfprepp#1;#2;#3 {{\par\rn#1 #2, #3\par}}

\def\beq#1{\begin{equation}\label{#1}}
\def\eeq{\end{equation}}
\def\beqa#1{\begin{eqnarray}\label{#1}}
\def\eeqa{\end{eqnarray}}

\def\spose#1{\hbox to 0pt{#1\hss}}
\def\simlt{\mathrel{\spose{\lower 3pt\hbox{$\mathchar"218$}}
     \raise 2.0pt\hbox{$\mathchar"13C$}}}
\def\simgt{\mathrel{\spose{\lower 3pt\hbox{$\mathchar"218$}}
     \raise 2.0pt\hbox{$\mathchar"13E$}}}
\def\simpropto{\mathrel{\spose{\lower 3pt\hbox{$\mathchar"218$}}
     \raise 2.0pt\hbox{$\propto$}}}

\def\ed{\end{document}}

\def\beq#1{\begin{equation}\label{#1}}
\def\eeq{\end{equation}}
\def\beqa#1{\begin{eqnarray}\label{#1}}
\def\eeqa{\end{eqnarray}}

\def\ignore#1{}

\def\simless{\mathbin{\lower 3pt\hbox
        {$\,\rlap{\raise 5pt\hbox{$\char'074$}}\mathchar"7218\,$}}} 
\def\simgreat{\mathbin{\lower 3pt\hbox
        {$\,\rlap{\raise 5pt\hbox{$\char'076$}}\mathchar"7218\,$}}} 

\documentclass[amsmath,nofootinbib]{revtex4} 
\usepackage{url}
\begin{document}
\input{epsf.sty}



\def\mit{1}
\def\tucson{2}
\def\pton{3}
\def\osu{4}
\def\nyu{5}
\def\chicago{6}
\def\fnal{7}
\def\tokyo{8}
\def\colorado{9}
\def\portsmouth{10}
\def\jadwin{11}
\def\lbnl{12}
\def\psu{13}
\def\ictp{14}
\def\hopkins{15}
\def\drexel{16}
\def\case{17}
\def\lanl{18}
\def\washington{19}
\def\saao{20}
\def\uct{21}
\def\apo{22}
\def\barcelona{23}
\def\pitt{24}
\def\tokyoastro{25}
\def\harvard{26}
\def\flagstaff{27}
\def\penn{28}
\def\jpl{29}
\def\caltech{30}
\def\efi{31}
\def\gatan{32}
\def\sussex{33}
\def\seoul{34}
\def\rochester{35}
\def\hawaii{36}

\def\affilmrk#1{$^{#1}$}
\def\affilmk#1#2{$^{#1}$#2;}


\title{Relativity Revisited}

\author{Flora Lopis \& Max Tegmark}

\address{Dept.~of Physics, Massachusetts Institute of Technology, Cambridge, MA 02139, USA}

\date{April 1, 2008, submitted to Physical Refuse}

\begin{abstract}
Was Einstein wrong? 
This paper provides a detailed technical review of Einstein's special and general relativity 
from an astrophysical perspective, 
including the historical development of the theories, experimental tests, 
modern applications to black holes, cosmology and parallel universes,
and last but not least, novel ways of expressing their seven most important equations.
\end{abstract}


  
\maketitle


\section{Introduction}
\label{IntroSec}

About a century after their inception, Einstein's theories of special \cite{Einstein05} and general \cite{Einstein16} relativity remain as 
topical as ever. The aim of this paper, developed for integration into the curriculum for MIT's 
relativity course 8.033, is to provide a detailed technical review of both the special and general theories
from an astrophysical perspective. Particular emphasis is placed on 
the historical development of the theories, experimental tests and modern applications such as 
black holes, cosmology, eternal inflation and parallel universes.
Novel ways of expressing the seven most important equations are presented, which is especially timely today \cite{Scott06,Scott07,Scott08,Follop07,Miralda07}.

To maximize the learning experience from this technical review, the reader is encouraged to sing it to 
the tune of {\it Yellow Submarine}, with italicized lines going like the chorus,
ideally to guitar accompaniment by Enectali Figueroa.\footnote{An example of this is provided at \url{http://www.youtube.com/watch?v=5PkLLXhONvQ}}

\section{Special Relativity}

\begin{center}

R\"omer measured the speed of light, \\
and something basic just wasn't right.\\ 
because Michaelson and Morley \\
showed that aether fit data poorly.\\ 

\bigskip

We jump to 1905. \\
In Einstein's brain, ideas thrive: \\
``The laws of nature must be the same \\
in every inertial frame'' \\
{\it We all believe in relativity, relativity, relativity.} \\
{\it Yes we all believe in relativity, 8.033, relativity.} \\
\bigskip

Einstein's postulates imply \\
that planes are shorter when they fly. \\
Their clocks are slowed by time dilation, \\
and look warped from aberration. \\
{\it Cos theta-prime is cos theta minus beta ... over one minus beta cos theta.} \\
{\it Yes we all believe in relativity, 8.033, relativity.} \\
\bigskip

With the Lorentz transformation, \\
we calculate the relation \\
between Chris's and Zoe's frame, \\
but all invariants, they are the same. \\
{\it Like $B$ dot $E$ and $B$-squared minus $E$-squared,} \\
{\it ... and the rest mass squared which is $E$-squared minus $p$-squared.} \\
{\it  'cos we all believe in relativity, 8.033, relativity.} \\

\clearpage

\bigskip
\begin{figure} 
\centerline{\epsfxsize=12cm\epsffile{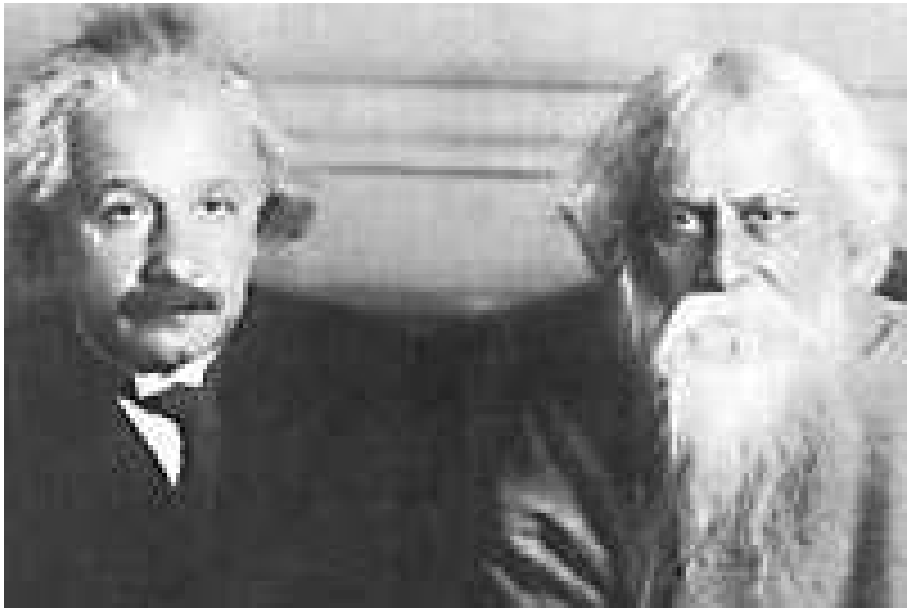}}
\caption[1]{
Einstein contemplating his never published theory of Tagorelativity.\\
(This image was taken by American photographer 
Martin Vos in 1930 and published in 1930 without copyright notice.)
}
\end{figure}

Soon physicists had a proclivity \\
for using relativity. \\
But nukes made us all scared \\
because $E=mc^2$. \\
{\it Everything is relative, even simultaneity,}\\ 
{\it and soon Einstein's become a de facto physics deity. }\\
{\it 'cos we all believe in relativity, 8.033, relativity.}\\
\end{center}

\section{General Relativity}

\begin{center}

But Einstein had another dream, \\
and in nineteen sixteen \\
he made a deep unification \\
between gravity and acceleration. \\
He said physics ain't hard at all \\
as long as you are in free fall, \\
'cos our laws all stay the same \\
in a locally inertial frame. \\
{\it And he called it general relativity, relativity, relativity.}\\
{\it And we all believe in relativity, 8.033, relativity.} \\
\bigskip

If towards a black hole you fall \\
tides will make you slim tall, \\
but your friends won't see you enter \\
a singularity at the center, \\
because it will look to them \\
like you got stuck at radius $2M$. \\
But you get squished, despite this balking, \\
and then evaporate, says Stephen Hawking. \\
{\it We all believe in relativity, relativity, relativity.} \\
{\it Yes we all believe in relativity, 8.033, relativity.} \\
\bigskip

\clearpage
We're in an expanding space \\
with galaxies all over the place, \\
and we've learned from Edwin Hubble \\
that twice the distance makes redshift double \\
We can with confidence converse \\
about the age of our universe. \\
Rival theories are now moot \\
thanks to Penzias, Wilson, Mather \& Smoot. \\
{\it We all live in an expanding universe, expanding universe, expanding universe.}\\
{\it Yes we all live in an expanding universe, expanding universe, expanding universe.}\\
\bigskip

But what's the physics of creation? \\
There's a theory called inflation \\
by Alan Guth and his friends, \\
but the catch is that it never ends, \\
making a fractal multiverse \\
which makes some of their colleagues curse. \\
Yes there's plenty left to figure out \\
like what reality is all about about. \\
{\it but at least we believe in relativity, relativity, relativity.}\\
{\it Yes we all believe in relativity, 8.033, relativity.}
\bigskip

\end{center}

\bigskip
{\bf Acknowledgments:}

\bigskip
We thank Ang\'elica de Oliveira-Costa, Betsy Devine, Enectali Figueroa and Frank Wilczek for helpful,
encouraging and disparaging comments on an earlier draft of the manuscript.
This work was unsupported by NASA grants NAG5-11099 and NNG06GC55G,
NSF grants AST-0134999 and 0607597, the Kavli Foundation, and fellowships from the David and Lucile
Packard Foundation and the Research Corporation. 



\begin{thebibliography}{99}


\bibitem{Einstein05}
\rf\nn Einstein A;1905;Annalen Der Physik;17;891

\bibitem{Einstein16}
\rf\nn Einstein A;1916;Annalen der Physik;49;284

\bibitem{Scott06}
\rfprep\nn Scott D\dualand\nn Frolop A;2006;astro-ph/0604011

\bibitem{Scott07}
\rfprep\nn Scott D\dualand\nn Frolop A;2007;astro-ph/0703783

\bibitem{Scott08}
\rfprep\nn Scott D\dualand\nn Frolop A;2008;{arXiv:0803.4378 [astro-ph]}

\bibitem{Follop07}
\rfprep\nn Follop R, \nn Rassat A, \nn Cooray A\multiand\nn Abdalla F;2007;astro-ph/0703806

\bibitem{Miralda07}
\rfprep\nn {Miralda-Escude} J;2007;astro-ph/0703774


\end{thebibliography}
\end{document}